\documentclass{ws-procs975x65}

\begin{document}

\title{NONCOMMUTATIVE QUANTUM MECHANICS VIEWED FROM FEYNMAN FORMALISM}

\author{J. LAGES}

\address{Institut UTINAM, Dynamique des Structures Complexes, UMR du CNRS
6213,\\
Universit\'e de Franche-Comt\'e, Campus La Bouloie, 25030 Besan\c
con Cedex, France\\ E-mail: jose.lages@utinam.cnrs.fr}

\author{A. BERARD, H. MOHRBACH and Y. GRANDATI}

\address{Laboratoire de Physique Mol\'eculaire et des Collisions,\\
 ICPMB, IF CNRS N$^\circ $ 2843, Universit\'e Paul Verlaine,\\
Institut de Physique, Bd Arago, 57078 Metz Cedex 3, France\\}

\author{P. GOSSELIN}

\address{Institut Fourier, Universit\'e de Grenoble I, UMR
5582 CNRS-UJF,\\
UFR de Math\'ematiques, BP74, 38402 St Martin d'H\`eres, Cedex
France}

\begin{abstract}
Dyson published in 1990 a proof due to Feynman of the Maxwell
equations. This proof is based on the assumption of simple
commutation relations between position and velocity. We first study
a nonrelativistic particle using Feynman formalism. We show that
Poincar\'{e}'s magnetic angular momentum and Dirac magnetic monopole
are the direct consequences of the structure of the sO(3) Lie
algebra in Feynman formalism. Then we show how to extend this
formalism to the dual momentum space with the aim of introducing
Noncommutative Quantum Mechanics which was recently the subject of a
wide range of works from particle physics to condensed matter
physics.
\end{abstract}

\keywords{Noncommutative Quantum Mechanics, Feynman formalism,
Dirac magnetic monopole, Poincar\'e momentum, Maxwell's equations,
Angular algebra symmetry, Dirac quantization, Berry curvature}

\bodymatter

\section{Introduction}

Feynman's ideas\cite{DYSON} were exposed by Dyson in an elegant
publication. Initial Feynman's motivation was to develop a
quantization procedure without resort to a Lagrangian or a
Hamiltonian. Assuming minimal commutation relations between
position and velocity and using Newton's second law, Feynman
derived the sourceless set of Maxwell's equations which are
galileo invariant. The interpretation of the Feynman's derivation
of the Maxwell's equations has aroused\cite
{TANIMURA,NOUS3,LEE,CHOU,CARINENA,HOJMAN,HUGHES,MONTESINOS,SINGH,SILAGADZE,BOUL}
a great interest among physicists. In particular Tanimura
\cite{TANIMURA} has generalized the Feynman's derivation in a
Lorentz covariant form with a scalar time evolution parameter. An
extension of the Tanimura's approach has been achieved\cite{NOUS3}
in using the Hodge duality to derive the two groups of Maxwell's
equations with a magnetic monopole in a f\/{}lat and in a curved
spaces. In Ref.~\refcite{LEE} the descriptions of relativistic and
non relativistic particles in an electromagnetic f\/ield was
studied, whereas in Ref.~\refcite{CHOU} a dynamical equation for
spinning particles was proposed. A rigorous mathematical
interpretation of Feynman's derivation connected to the inverse
problem for the Poisson dynamic has been formulated in
Ref.~\refcite {CARINENA}. Also in Refs.~\refcite{HOJMAN} and
\refcite{HUGHES} the Feynman's derivation is considered in the
frame of the Helmholtz's inverse problem for the calculus of
variations. Other works\cite{MONTESINOS,SINGH,SILAGADZE} have
provided new looks on the Feynman's derivation of the Maxwell's
equations. More recently\cite{NOUS6}, some of the authors embedded
Feynman's derivation of the Maxwell's equation in the framework of
noncommutative geometry. As Feynman's brackets can be interpreted
as a deformation of Poisson brackets we showed that the Feynman
brackets can be viewed as a generalization of the Moyal brackets
defined over the tangent bundle space\cite{NOUS6}.

The mathematical foundations of Feynman's formalism is presented
in Section \ref{formalism} and is used to review the Feynman's
derivation of the Maxwell's equation in Section \ref{maxwell}. It
is well known that velocities do not commute in the presence of an
electromagnetic field. For this reason the angular algebra
symmetry, \emph{e.g.} the sO(3) symmetry in the Euclidean case, is
broken. In Section \ref{so3} we show how to restore such a
symmetry and we point out in this context the necessity of adding
the Poincar\'{e} momentum $\mathbf{M}$ to the simple angular
momentum $\mathbf{L} $. The direct consequence of this restoration
is then the generation of a Dirac magnetic monopole. A natural
extension of Feynman's formalism is to consider the dual momentum
space. In Section \ref{NQM} we embed then our work in the natural
generalization of Quantum Mechanics involving noncommutative
coordinates. This generalization was originally introduced by
Snyder\cite{SNYDER} as a short distance regularization to improve
the problem of infinite self energies inherent in a Quantum Field
Theory. Due to the advent of the renormalization theory this idea
was not very popular until Connes\cite{CONNES} analyzed Yang Mills
theories on noncommutative space. More recently a correspondence
between a noncommutative gauge theory and a conventional gauge
theory was introduced by Seiberg and Witten\cite{SEIBERG}.
Noncommutative gauge theories were also found as being naturally
related to string and M-theory\cite{KONECHNY}. Applications of
noncommutative theories were also found in condensed matter
physics, for instance in the Quantum Hall effect \cite{BELLISSARD}
and in the noncommutative Landau problem \cite{JACKIW1,HORVATHY},
the name of Noncommutative Quantum Mechanics started then to be
used\cite{HORVATHY,GAMBOA,NAIR,CORTESE}. In section \ref{NQM} we resume
our paper \cite{NOUS7} which shows that in our model a quantum
particle in a harmonic potential has a behavior similar to a
particle in a constant magnetic field $\theta $ in standard
quantum mechanics, since a paramagnetic term appears in the
Hamiltonian. Moreover the particle acquires an effective dual mass
in the same way that an electron moving in a periodic potential in
solid state physics. Again the angular algebra symmetry is
naturally broken and the restoration of this symmetry gives then a
dual Dirac monopole in momentum space field configuration.

\section{Mathematical foundations of Feynman's formalism}

\label{formalism}

Let a particle with a mass $m$ and an electrical charge $q$ be
described by the vector $\mathbf{x}=\{x^{i}\}_{i=1,\dots ,N}$ which
def\/{}ines its
position on the manifold $\mathcal{M}$. Let the manifold $\mathcal{M}$ be a $%
N$-dimensional vectorial manifold diffeomorphic to $\Bbb{R}^{N}$.
Let $\tau $ be the parameter of the group of diffeomorphisms
$\mathcal{G}:\Bbb{R}\times
\mathcal{M}\rightarrow \mathcal{M}$ such as $\mathcal{G}(\tau ,\mathbf{x})=%
\mathcal{G}^{\tau }\mathbf{x}=\mathbf{x}(\tau )$. Then taking $\tau
$ as the time parameter of our physical system we are able to
def\/{}ine a velocity
vector $\dot{\mathbf{x}}\in \mathcal{M}$ as $\dot{\mathbf{x}}=\displaystyle%
\frac{d\mathbf{x}}{d\tau }=\mathcal{G}^{\tau
}\mathbf{x}=\{\dot{x}^{i}(\tau )\}_{i=1,\dots ,N}$. Let
$T(\mathcal{M})$ be the tangent bundle space associated with the
manifold $\mathcal{M}$, a point on $T(\mathcal{M})$ is
described then by a $2N$ dimensional vector $\mathbf{\boldsymbol{\xi}}=\{\mathbf{x},\mathbf{%
\dot{x}}\}$. Let $A^{0}(T(\mathcal{M}))=C^{\infty
}(T(\mathcal{M}),\Bbb{R})$
be the algebra of differential functions def\/{}ined on the manifold $T(%
\mathcal{M}) $. We def\/{}ine a Poisson structure on
$T(\mathcal{M})$ which
is an internal skew-symmetric bilinear multiplicative law on $A^{0}(T(%
\mathcal{M})) $ denoted $(f,g)\rightarrow \lbrack f,g]$ and
satisfying the Leibnitz rule
\begin{equation}
\lbrack f,gh]=[f,g]h+[f,h]g  \label{Leibnitz}
\end{equation}
and the Jacobi identity
\begin{equation}
J(f,g,h)=[f,[g,h]]+[g,[h,f]]+[h,[f,g]]=0.  \label{Jacobi}
\end{equation}
The manifold $T(\mathcal{M})$ with such a Poisson structure is
called a
Poisson manifold. We def\/{}ine a dynamical system on the Poisson manifold $%
T(\mathcal{M})$ by the following differential equation
\begin{equation}
\frac{df}{d\tau }=[f,H]  \label{dynamic}
\end{equation}
where $H\in A^{0}(T(\mathcal{M}))$ is the Hamiltonian of the
dynamical system.

With such def\/{}initions we derive the following important
relations for functions belonging to $A^{0}(T(\mathcal{M}))$
\begin{eqnarray}
\left[ f(\mathbf{\boldsymbol{\xi}}),h(\mathbf{\boldsymbol{\xi}})\right] &=&\left\{ f(\mathbf{\boldsymbol{\xi}}),h(\mathbf{%
\boldsymbol{\xi}})\right\}  \nonumber \\
&+&\left[ x^{i},x^{j}\right] \frac{\partial f(\mathbf{\boldsymbol{\xi}})}{\partial x^{i}}%
\frac{\partial h(\mathbf{\boldsymbol{\xi}})}{\partial x^{j}}+\left[ \dot{x}^{i},\dot{x}%
^{j}\right] \frac{\partial f(\mathbf{\boldsymbol{\xi}})}{\partial \dot{x}^{i}}\frac{%
\partial h(\mathbf{\boldsymbol{\xi}})}{\partial \dot{x}^{j}},  \label{bracket1}
\end{eqnarray}
where we have introduced Poisson-like brackets def\/{}ined by
\begin{equation}
\left\{
f(\mathbf{\boldsymbol{\xi}}),g(\mathbf{\boldsymbol{\xi}})\right\}
=\left[ x^{i},\dot{x}^{j}\right]
\left( \frac{\partial f(\mathbf{\boldsymbol{\xi}})}{\partial x^{i}}\frac{\partial h(\mathbf{%
\boldsymbol{\xi}})}{\partial \dot{x}^{j}}-\frac{\partial f(\mathbf{\boldsymbol{\xi}})}{\partial \dot{x}^{i}%
}\frac{\partial h(\mathbf{\boldsymbol{\xi}})}{\partial x^{j}}\right)
. \label{Poissonlike}
\end{equation}
We can see the relation (\ref{bracket1}) as the simple deformation
of the Poisson-like brackets introduced in (\ref{Poissonlike}). It
is obvious that
the tensors $\left[ x^{i},x^{j}\right] $ and $\left[ \dot{x}^{i},\dot{x}%
^{j}\right] $ are skew symmetric.. We introduce then the following
notations
\[
\left[ x^{i},x^{j}\right] =\frac{q}{m^{2}}\;\theta ^{ij}(\mathbf{\boldsymbol{\xi}}%
),\;\;\;\left[ x^{i},\dot{x}^{j}\right] =\frac{1}{m}\;g^{ij}(\mathbf{\boldsymbol{\xi}}%
),\;\;\;\left[ \dot{x}^{i},\dot{x}^{j}\right] =\frac{q}{m^{2}}\;F{}^{ij}(%
\mathbf{\boldsymbol{\xi}})
\]
where $g^{ij}(\mathbf{\boldsymbol{\xi}})$ is the $N\times N$ metric tensor, and where $%
\theta ^{ij}(\mathbf{\boldsymbol{\xi}})$ and
$F^{ij}(\mathbf{\boldsymbol{\xi}})$ are two $N\times N$ skew
symmetric tensors, $F^{ij}(\mathbf{\boldsymbol{\xi}})$ being related
to the electromagnetic tensor introduced in a preceding paper
\cite{NOUS1}.

\section{Maxwell's equations}

\label{maxwell}

In a three dimensional f\/{}lat space we have
$g^{ij}(\mathbf{x})=\delta ^{ij}$ and the Hamiltonian of the Poisson
structure reads then
\begin{equation}
H=\frac{1}{2}m\dot{x}^{i}\dot{x}_{i}+f(\mathbf{x}).
\end{equation}
The Jacobi identity (\ref{Jacobi}) involving position and velocity
components
\begin{equation}
\displaystyle J(x^{i},\dot{x}^{j},\dot{x}^{k})\propto \frac{\partial F^{jk}(%
\mathbf{\boldsymbol{\xi}})}{\partial \dot{x}^{i}}=0.
\end{equation}
shows that the gauge curvature is velocity independent, $F^{ij}(\mathbf{\boldsymbol{\xi}}%
)\equiv F^{ij}(\mathbf{x})$. From the Jacobi identity (\ref{Jacobi})
involving only velocities components we derive the Bianchi equation
\begin{equation}
\displaystyle J(\dot{x}^{i},\dot{x}^{j},\dot{x}^{k})\propto
\varepsilon ^{k}{}_{ji}\displaystyle\frac{\partial
F^{ij}(\mathbf{x})}{\partial x^{k}}=0 \label{Jacobi1}
\end{equation}
which, if we set $F^{ij}(\mathbf{x})=\varepsilon ^{ji}{}_{k}B^{k}(\mathbf{x}%
) $, gives the following Maxwell's equation
\begin{equation}
\mathbf{\nabla }\cdot \mathbf{B}=0.  \label{divB}
\end{equation}
Now using the dynamical equation (\ref{dynamic}) we obtain the
following equation of motion
\begin{equation}
m\ddot{x}^{i}=m\left[ \dot{x}^{i},H\right] =qF^{ij}(\mathbf{x})\dot{x}%
_{j}+qE^{i}\left( \mathbf{x}\right)  \label{eqmotion}
\end{equation}
where
\begin{equation}
qE^{i}\left( \mathbf{x}\right) =-\frac{\partial f(\mathbf{x})}{\partial x_{i}%
}.  \label{electro}
\end{equation}
We have then a particle of mass $m$ and electrical charge $q$ moving
in f\/{}lat space where a magnetostatic and an electrostatic
external f\/ield are present. We are able now to derive the other
Maxwell's equation of the f\/{}irst group. With the dynamical
equation (\ref{dynamic}) we express the time derivative of the
magnetic f\/ield
\begin{equation}
\displaystyle\frac{dB^{i}}{dt}=\displaystyle\frac{1}{2}\varepsilon
^{i}{}_{jk}\left[ F^{jk},H\right]
=\displaystyle\frac{m^{2}}{2q}\varepsilon ^{i}{}_{jk}\left[ \left[
\dot{x}^{j},\dot{x}^{k}\right] ,H\right]
\end{equation}
and we use the Jacobi identy (\ref{Jacobi}) to rewrite the last term
of the last equation. After some calculus we obtain
\begin{equation}
\displaystyle\frac{dB^{i}}{dt}=-\dot{x}^{i}\mathbf{\nabla }\cdot \mathbf{B}+%
\displaystyle\frac{\partial B^{i}}{\partial
x_{j}}\dot{x}_{j}+\varepsilon ^{i}{}_{jk}\displaystyle\frac{\partial
E^{j}}{\partial x_{k}}
\end{equation}
which using (\ref{divB}) gives the second Maxwell's equation
\begin{equation}
\displaystyle\frac{\partial \mathbf{B}}{\partial t}-\mathbf{\nabla
}\times \mathbf{E}=\mathbf{0}
\end{equation}
for static f\/ields and electric f\/ields deriving from any potential $f(%
\mathbf{x})$ (\ref{electro}).

As the two other Maxwell's equations are not Galileo invariant they
cannot be deduced from the formalism and can be merely seen as a
def\/{}inition of the charge density and the current density.
Nevertheless, as shown in Ref.~\refcite{NOUS6} the complete set of
the Maxwell's equations can be deduced in the relativistic
generalization.

\section{sO(3) algebra and Poincar\'e momentum}

\label{so3}

One of the most important symmetry in physics is naturally the
spherical symmetry corresponding to the isotropy of the physical
space. This symmetry is related to the sO(3) algebra. In the
following we show that this symmetry is broken when an
electromagnetic f\/ield is applied. In order to study the
symmetry breaking of the sO(3) algebra we use the usual angular momentum $%
L^{i}=m\varepsilon ^{i}{}_{jk}x^{j}\dot{x}^{k}$ which is a constant
of motion in absence of gauge f\/ield. In fact, no electromagnetic
f\/ield implies $F^{ij}(\mathbf{x})=\left[
\dot{x}^{i},\dot{x}^{j}\right] =0$, and the expression of the sO(3)
Lie algebra with our brackets (\ref{bracket1}) gives then the
standard algebra def\/{}ined in terms of the Poisson brackets
(\ref{Poissonlike})
\begin{equation}
\left\{
\begin{array}{l}
\left[ x^{i},L^{j}\right] =\left\{ x^{i},L^{j}\right\} =\varepsilon
^{ij}{}_{k}x^{k}, \\
\left[ \dot{x}^{i},L^{j}\right] =\left\{ \dot{x}^{i},L^{j}\right\}
=\varepsilon ^{ij}{}_{k}\dot{x}^{k}, \\
\left[ L^{i},L^{j}\right] =\left\{ L^{i},L^{j}\right\} =\varepsilon
^{ij}{}_{k}L^{k}.
\end{array}
\right.
\end{equation}
When the electromagnetic f\/ield is turned on this algebra is broken
in the following manner
\begin{equation}
\left\{
\begin{array}{ccccl}
\left[ x^{i},L^{j}\right] & = & \left\{ x^{i},L^{j}\right\} & = &
\varepsilon ^{ij}{}_{k}x^{k}, \\
\left[ \dot{x}^{i},L^{j}\right] & = & \left\{
\dot{x}^{i},L^{j}\right\} & +
& \frac{q}{m}\varepsilon ^{j}{}_{kl}x^{k}{}F^{il}(\mathbf{x}) \\
& = & \varepsilon ^{ij}{}_{k}\dot{x}^{k} & + &
\frac{q}{m}\varepsilon
^{j}{}_{kl}x^{k}{}F^{il}(\mathbf{x}), \\
\left[ L^{i},L^{j}\right] & = & \left\{ L^{i},L^{j}\right\} & + &
q\varepsilon ^{i}{}_{kl}\varepsilon
^{j}{}_{ms}x^{k}x^{m}{}F^{ls}(\mathbf{x})
\\
& = & \varepsilon ^{ij}{}_{k}L^{k} & + & q\varepsilon
^{i}{}_{kl}\varepsilon ^{j}{}_{ms}x^{k}x^{m}{}F^{ls}(\mathbf{x}).
\end{array}
\right.
\end{equation}
In order to restore the sO(3) algebra we introduce a new angular momentum $%
M^{i}(\mathbf{\boldsymbol{\xi}})$ which is \textit{a priori}
position and velocity dependent. We consider then the following
transformation law
\begin{equation}
L^{i}(\mathbf{\boldsymbol{\xi}})\rightarrow \mathcal{L}^{i}(\mathbf{\boldsymbol{\xi}})=L^{i}(\mathbf{\boldsymbol{\xi}}%
)+M^{i}(\mathbf{\boldsymbol{\xi}}),
\end{equation}
and we require that this new angular momentum $\mathcal{L}^{i}$
verif\/{}ies the usual sO(3) algebra
\begin{equation}
\left\{
\begin{array}{l}
\left[ x^{i},\mathcal{L}{}^{j}\right] =\left\{
x^{i},\mathcal{L}^{j}\right\}
=\varepsilon ^{ij}{}_{k}x^{k}, \\
\left[ \dot{x}^{i},{}\mathcal{L}^{j}\right] =\left\{ \dot{x}^{i},\mathcal{L}%
{}^{j}\right\} =\varepsilon ^{ij}{}_{k}\dot{x}^{k}, \\
\left[ {}\mathcal{L}^{i},{}\mathcal{L}^{j}\right] =\left\{ \mathcal{L}{}^{i},%
\mathcal{L}{}^{j}\right\} =\varepsilon ^{ij}{}_{k}\mathcal{L}^{k}.
\end{array}
\right.  \label{structure}
\end{equation}
These equations (\ref{structure}) gives then three constrains on the
expression of the angular momentum $\mathcal{L}^{i}$. From the
f\/{}irst relation in (\ref{structure}) we easily deduce that
$M^{i}$ is velocity independent
$
M{}^{i}(\mathbf{\boldsymbol{\xi}})=M^{i}(\mathbf{x}),
$
from the second relation we obtain
\begin{equation}
\left[ \dot{x}^{i},M^{j}\right] =-\frac{1}{m}\frac{\partial M^{j}(\mathbf{x})%
}{\partial x_{i}}=-\frac{q}{m}\varepsilon
^{j}{}_{kl}x^{k}F^{il}(\mathbf{x}) \label{momentum1}
\end{equation}
and f\/{}inally the third relation gives
\begin{equation}
M^{i}=\frac{1}{2}q\varepsilon
_{jkl}x^{i}x^{k}F^{jl}(\mathbf{x})=-q\left( \mathbf{x}\cdot
\mathbf{B}\right) x^{i}.  \label{momentum}
\end{equation}
Equations (\ref{momentum1}) and (\ref{momentum}) are compatible only
if the magnetic f\/ield $\mathbf{B}$ is the Dirac magnetic monopole
f\/ield
\begin{equation}
\mathbf{B}=\displaystyle\frac{g}{4\pi }\frac{\mathbf{x}}{\left\| \mathbf{x}%
\right\| ^{3}}.
\end{equation}
The vector $\mathbf{M}$ allowing us to restore the sO(3) symmetry
(\ref {structure}) is then the Poincar\'{e} momentum \cite{POINCARE}
\[
\mathbf{M}=-\frac{qg}{4\pi }\frac{\mathbf{x}}{\left\|
\mathbf{x}\right\| }.
\]
already found in a preceding paper \cite{NOUS4}\cite{NOUS2}. The
total angular momentum is then
\begin{equation}
\boldsymbol{\mathcal{L}}=\mathbf{L}-\frac{qg}{4\pi
}\frac{\mathbf{x}}{\left\| \mathbf{x}\right\| }.
\end{equation}
This expression was initially found by Poincar\'{e} in a different
context \cite{POINCARE}. Actually he was looking for a new angular
momentum that would be a constant of motion. In our framework this
property is trivially verif\/{}ied by using the dynamical relation
(\ref{dynamic}). This procedure of symmetry restoration has also
been performed for Lorentz algebra in a curved space\cite{NOUS6}.
An other generalization of this formalism can be find in a recent
interesting work where the study of the Lorentz generators in
N-dimensional Minkowski space is proposed \cite{LAND1,LAND2}.

Let us now discuss an important point. As the Dirac magnetic
monopole is located at the origin we have
\begin{equation}
J(\dot{x}^{i},\dot{x}^{j},\dot{x}^{k})=\mathbf{\nabla }\cdot \mathbf{B}%
=g\delta ^{3}(\mathbf{x}).  \label{xpxpxpjackiw}
\end{equation}
The preservation of the sO(3) symmetry in the presence of a gauge
f\/ield is then incompatible with the requirement of the Jacobi
identity at the origin
of the coordinates and we have to exclude the origin from the manifold $%
\mathcal{M}$. As the Jacobi identity is the inf\/{}initesimal
statement of associativity in the composition law of the translation
group \cite{JACKIW},
the breakdown of the Jacobi identity (\ref{xpxpxpjackiw}) when $\mathbf{%
\nabla }\cdot \mathbf{B}\neq 0$ implies that f\/{}inite translations
do not associate. In usual quantum mechanics non-associativity
between operators acting on the Hilbert space can not be tolerate,
one has to use the Dirac's quantization procedure to save
associativity (\ref{xpxpxpjackiw}).


In order to consider quantum mechanics within our framework we have
to quantify as usual the total angular momentum $\mathcal{L}$.
Considering the rest frame of the particle we have the following
Dirac quantization
\begin{equation}
\frac{gq}{4\pi }=\frac{n}{2}\hbar .
\end{equation}

\section{Noncommutative Quantum Mechanics}

\label{NQM}

Let now the momentum vector $\mathbf{p}$ replace the velocity vector $\dot{%
\mathbf{x}}$ in the Feynman formalism presented in Sec.
\ref{formalism}. Consider a quantum particle of mass $m$ whose
coordinates satisfy the deformed Heisenberg algebra
\[
\left[ x^{i},x^{j}\right] =i\hbar q_{\theta }\theta ^{ij}(\mathbf{x},\mathbf{%
\ p}),\;\;\;\left[ x^{i},p^{j}\right] =i\hbar \delta
^{ij},\;\;\;\left[ p^{i},p^{j}\right] =0,
\]
where $\theta $ is a field which is \emph{a priori} position and
momentum dependent and $q_{\theta }$ is a charge characterizing the
intensity of the interaction of the particle and the $\theta $
field. The commutation of the momentum implies that there is no
external magnetic field. It is well known that these commutation
relations can be obtained from the deformation of the Poisson
algebra of classical observable with a provided Weyl-Wigner-Moyal
product \cite{MOYAL} expanded at the first order in $\theta $.

\subsection{Jacobi identities}

The Jacobi identity $J(p^{i},x^{j},x^{k})=0$ implies the important
property
that the $\theta $ field is position independent $\theta ^{jk}=\theta ^{jk}(%
\mathbf{p})$. Then one can see the $\theta $ field like a dual of a
magnetic field and $q_{\theta }$ like a dual of an electric charge.
The fact that the field is homogeneous in space is an essential
property for the vacuum. In addition, one easily see that a particle
in this field moves freely, that is, the vacuum field does not act
on the motion of the particle in the absence of an external
potential. The effect of the $\theta $ field is manifest only in
presence of a position dependent potential. To look further at the
properties of the $\theta $ field consider the other Jacobi identity
$J(x^{i},x^{j},x^{k})=0$ giving the equation of motion of the field
\begin{equation}
\frac{\partial \theta ^{jk}(\mathbf{p})}{\partial
p^{i}}+\frac{\partial
\theta ^{ki}(\mathbf{p})}{\partial p^{j}}+\frac{\partial \theta ^{ij}(%
\mathbf{p})}{\partial p^{k}}=0,  \label{div}
\end{equation}
which is the dual equation of the Maxwell equation $\mathbf{\nabla
}\cdot \mathbf{B}=0$. As we will see later, equation (\ref{div}) is
not satisfied in the presence of a monopole and this will have
important consequences.

\subsection{Position transformation}

Now consider the position transformation $X^{i}=x^{i}+q_{\theta
}a_{\theta }^{i}(\mathbf{x},\mathbf{p})$, where $a_{\theta }$ is
\emph{a priori} position and momentum dependent, that restores the
usual canonical Heisenberg algebra
\[
\left[ X^{i},X^{j}\right] =0,\;\;\;\left[ X^{i},p^{j}\right] =i\hbar
\delta ^{ij},\;\;\;\left[ p^{i},p^{j}\right] =0.
\]
The second commutation relation implies that $a_{\theta }$ is
position independent, while the commutation relation of the
positions leads to the
following expression of $\theta $ in terms of the dual gauge field $%
a_{\theta }$
\begin{equation}
\theta ^{ij}(\mathbf{p})=\frac{\partial a_{\theta }^{i}(\mathbf{p})}{%
\partial p^{j}}-\frac{\partial a_{\theta }^{j}(\mathbf{p})}{\partial p^{i}},
\label{theta}
\end{equation}
which is dual to the standard electromagnetic relation in position
space.

\subsection{Field Properties}

In order to examine more in detail the properties of this new field,
let us consider initially the case of a constant field what is usual
in noncommutative quantum mechanics. In the case of an harmonic
oscillator expressed in terms of the original
$\{\mathbf{x},\mathbf{p}\}$ coordinates the Hamiltonian reads
\begin{equation}
H_{\theta }(\mathbf{x},\mathbf{p})=\frac{\mathbf{p}^{2}}{2}+\frac{k}{2}%
\mathbf{x}^{2}  \label{hamilton}
\end{equation}
from which we have $p^{i}=m\dot{x}^{i}-kq_{\theta }\theta ^{ij}x_{j}$, $\dot{%
p}^{i}=-kx^{i}$\ and the equation of motion $
m\ddot{x}^{i}=kq_{\theta }\theta ^{ij}\dot{x}_{j}-kx^{i}$ which
corresponds formally to a particle in a harmonic oscillator
submitted to an external constant magnetic field. From equation
(\ref{theta}) we deduce that $a_{\theta }^{i}(\mathbf{p})=q_{\theta
}$\ $\theta ^{ij}p_{j}$, so $X^{i}=x^{i}+\frac{1}{2}q_{\theta
}\theta ^{ij}p_{j}$, and the Hamiltonian can then be written
\begin{equation}
H_{\theta }(\mathbf{X},\mathbf{p})=\frac{\left( m_{*}^{-1}\right)
^{ij}p_{i}p_{j}}{2}+\frac{k}{2}\mathbf{X}^{2}-k\frac{q_{\theta }}{2m}\mathbf{%
\Theta }\cdot {\mathcal{L}},
\end{equation}
with $\theta ^{ij}=\varepsilon ^{ijk}\Theta _{k}$, $\mathcal{L}^{i}(\mathbf{X%
},\mathbf{p})=\frac{1}{2}\varepsilon ^{i}{}_{jk}\left(
X^{j}p^{k}+p^{k}X^{j}\right) $ and $\sigma ^{ij}=\delta ^{ij}\mathbf{\Theta }%
^{2}-\Theta ^{i}\Theta ^{j}$, the dual tensor of the Maxwell
constraint tensor. Note that the interaction with the field $\theta
$ is due to the presence of the position dependent harmonic
potential and leads to a dual paramagnetic interaction which could
be experimentally observable. Like an
electron in the effective periodic potential of ions, the particle in the $%
\theta $ field acquires an effective mass tensor $m_{*}^{ij}=m\left(
\delta ^{ij}+\frac{\hbar ^{2}kq_{\theta }^{2}}{4}\sigma ^{ij}\right)
^{-1}$ which breaks the homogeneity of space. This strong analogy
with the vacuum of the solid state leads us to consider this field
as a property of the vacuum.

\subsection{Angular momentum}

Consider now the problem of angular momentum. It is obvious that the
angular momentum expressed according to the canonical coordinates
satisfies the angular momentum algebra however it is not conserved
\begin{equation}
\frac{d}{dt}\boldsymbol{\mathcal{L}}(\mathbf{X},\mathbf{p})=kq_{\theta
} \boldsymbol{\mathcal{L}}\wedge \boldsymbol{\Theta }.
\end{equation}
In the original $\left\{x,p\right\} $ space the usual angular
momentum $\ L^{i}(\mathbf{x},\mathbf{p})=\varepsilon
^{i}{}_{jk}x^{j}p^{k}$, does not satisfy this algebra. So it seems
that there are no rotation generators in the $\left\{x,p\right\}$
space. We will now prove that a true angular momentum can be defined
only if $\theta $ is a non constant field. From the definition of
the angular momentum we deduce the following commutation relations
\begin{eqnarray}
&[x^{i},L^{j}]=i\hbar \varepsilon ^{ijk}x_{k}+i\hbar q_{\theta
}\varepsilon ^{j}{}_{kl}p^{l}\theta ^{ik}(\mathbf{p}),\;\;\;
[p^{i},L^{j}]=i\hbar \varepsilon ^{ijk}p_{k},\\
&[L^{i},L^{j}]=i\hbar
\varepsilon ^{ij}{}_{k}L^{k}+i\hbar q_{\theta }\varepsilon
^{i}{}_{kl}\varepsilon ^{j}{}_{mn}p^{l}p^{n}\theta ^{km}(
\mathbf{p}),
\end{eqnarray}
showing in particular that the sO(3) Algebra is broken. To restore
the angular momentum algebra consider the transformation law
\begin{equation}
L^{i}\rightarrow \Bbb{L}^{i}=L^{i}+M_{\theta }^{i}(\mathbf{x,p}),
\end{equation}
and require the usual algebra
\begin{equation}
[x^{i},\Bbb{L}^{j}]=i\hbar \varepsilon ^{ijk}x_{k},\;\;\;
[p^{i},\Bbb{L}^{j}]=i\hbar \varepsilon ^{ijk}p_{k},\;\;\;
[\Bbb{L}^{i},\Bbb{L}^{j}]=i\hbar \varepsilon ^{ijk}\Bbb{L}_{k}.
\label{lie}
\end{equation}
The second equation in (\ref{lie}) implies the position independent
property $M_{\theta }^{j}(\mathbf{x,p})=M_{\theta
}^{j}(\mathbf{p})$, while the third equation leads to
\begin{equation}
M_{\theta }^{i}(\mathbf{p})=\frac{1}{2}q_{\theta }\varepsilon
{}_{jkl}p^{i}p^{l}\theta ^{kj}(\mathbf{p}).
\end{equation}
Using this equation we rewrite the third equation in (\ref{lie}) and
we obtain a dual Dirac monopole\cite{DIRAC} defined in momentum
space
\begin{equation}
\mathbf{\Theta }(\mathbf{p})=\frac{g_{\theta }}{4\pi }\frac{%
\mathbf{p}}{\left\|\mathbf{p}\right\|^{3}}.  \label{monopole}
\end{equation}
We have introduced the dual magnetic charge $g_{\theta}$ associated
to the $\Theta$ field. Consequently we have
\begin{equation}
\mathbf{M}_{\theta}(\mathbf{p})=-\frac{q_{\theta }g_{\theta }}{4\pi
}\frac{\mathbf{p}}{\left\|\mathbf{p}\right\|}
\end{equation}
which is the dual of the famous Poincar\'e momentum introduced in
positions space \cite{POINCARE,NOUS1}. Then the generalized angular
momentum
\begin{equation}
\boldsymbol{\Bbb{L}}=\mathbf{r}\wedge \mathbf{p}-\frac{q_{\theta }g_{\theta }}{4\pi }\frac{\mathbf{p}}{%
\left\|\mathbf{p}\right\|}.
\end{equation}
is a genuine angular momentum satisfying the usual algebra. It is
the summation of the angular momentum of the particle and of the
dual monopole field. One can check that it is a conserved quantity.
It is interesting to note that the use of this formalism with these coordinates
had been made also for massless particles in the context of anyons \cite{CORTES}.

The duality between the monopole in momentum space and the Dirac
monopole is due to the symmetry of the commutation relations in
noncommutative quantum mechanics where $\left[ x^{i},x^{j}\right]
=i\hbar q_{\theta }\varepsilon ^{ijk}\Theta _{k}(\mathbf{p})$ and
the usual quantum mechanics in a magnetic field where $\left[
v^{i},v^{j}\right] =i\hbar q\varepsilon ^{ijk}B_{k}( \mathbf{x})$.
Therefore the two gauge fields $\Theta (\mathbf{p})$ and $B(
\mathbf{x})$ are dual to each other. Note that in the presence of \
the dual monopole the Jacobi identity fails
\begin{equation}
J(p^{i},x^{j},x^{k})=-q_{\theta }\hbar ^{2}\frac{\partial \Theta ^{i}(\mathbf{p})}{\partial p_{i}%
}=-4\pi q_{\theta }\hbar ^{2}g_{\theta }\delta ^{3}(\mathbf{p})\neq
0.
\end{equation}
This term is responsible for the violation of the associativity
which is only restored if the following quantification equation is
satisfied
\begin{equation}
\int d^{3}p\frac{\partial \Theta ^{i}}{\partial p_{i}}=\frac{2\pi n\hbar }{%
q_{\theta }}
\end{equation}
leading to $q_{\theta }g_{\theta }=\frac{n\hbar }{2},$ in complete
analogy with Dirac's quantization\cite{JACKIW}.

\subsection{Physical realization}

A physical realization of our theory was found very recently in the
context of the anomalous Hall effect in a ferromagnetic crystal
\cite{FANG}. The main point is the consideration of the Berry phase
\cite{BERRY} $ a_{n}^{\mu }(\mathbf{k})=i\left\langle
u_{n\mathbf{k}}\right| d_{k}\left| u_{n\mathbf{k}}\right\rangle $
where the wave function $u_{n\mathbf{k}}$ are the periodic part of
the Bloch waves. In their work, the authors introduced a gauge
covariant position operator of the wave packet associated to an
electron in the $n$ band $ x^{\mu }=i\frac{\partial }{\partial
k_{\mu }}-a_{n}^{\mu }(\mathbf{k})$, whose commutator is given by
\begin{equation}
\left[ x^{\mu },x^{\nu }\right] =\frac{\partial a_{n}^{\nu }(\mathbf{k})}{%
\partial k^{\mu }}-\frac{\partial a_{n}^{\mu }(\mathbf{k})}{\partial k^{\nu }%
}=-iF^{\mu \nu }(\mathbf{k})  \label{comm}
\end{equation}
where
$F^{\mu \nu }(\mathbf{k})$ is the Berry curvature in momentum space.

The connection with our noncommutative quantum mechanics theory is
then clearly apparent. The $\theta (\mathbf{p})$ field corresponds
to the Berry curvature $F(k)$ and $a_{\theta }(\mathbf{p})$ is
associated to the Berry phase $a_{n}(k)$. This shows that physical
situations with a Berry phase living in momentum space could be
expressed in the context of a noncommutative quantum mechanics. It
is essential to mention that the monopole in momentum space, that we
deduced from general symmetry considerations applied to the
noncommutative quantum mechanics was highlighted in very beautiful
experiments of Fang \emph{et al.} \cite{FANG}.

\section{Conclusion}

\label{conclusion}

Starting from the derivation of Maxwell's equations we reviewed
the Feynman formalism. The angular algebra symmetry is naturally
broken in the presence of a magnetic field, we showed within the
framework of the Feynman formalism how to restore this symmetry.
The restoration generates then a Dirac magnetic monopole and
implies in addition to the usual angular momentum an associated
Poincar\'e momentum. In Ref.~\refcite{NOUS6} this restoration has
been performed also in curved space and a direct application to
gravitoelectromagnetism has been given.

Going from the tangent bundle space to the cotangent bundle space
and requiring the restoration of the Heisenberg algebra, we have
shown that Noncommutative Quantum Mechanics can be viewed from
Feynman formalism. In order to maintain the sO(3) algebra a dual
monopole in momentum space is generated. This monopole is
responsible for the violation of the Jacobi identity and implies
the non associativity of the law of addition of the momentum. To
restore associativity a Dirac's quantization of the dual charges
is necessary. As a natural physical realization of our theory we
can see the $\theta(\mathbf{p})$ field like a Berry curvature
associated to a Berry phase expressed in momentum space. The
monopole in momentum space predicted by our generalization of
noncommutative quantum mechanics was found recently\cite{FANG} in
condensed matter physics experiments.


\end{document}